# Conceptual Modeling of Objects

Sabah Al-Fedaghi[*]

*Computer Engineering Department*
*Kuwait University*
*Kuwait*

salfedaghi@yahoo.com, sabah.alfedaghi@ku.edu.kw

*Abstract* – In this paper, we concentrate on object-related analysis in the field of general ontology of reality as related to software engineering (e.g., UML classes). Such a venture is similar to many studies in which researchers have enhanced modeling through ontological analysis of the underlying paradigm of UML models. We attempt to develop a conceptual model that consists of a foundation of *things* that is supplemented with a second level of designated *objects*. According to some researchers, the problem of the difference between things and objects is one of the most decisive issues for the conception of reality. In software engineering, objects serve two purposes: they promote understanding of the real world and provide a practical basis for computer implementation. The notion of *object* plays a central role in the object-oriented approach, in which other notions are viewed by decomposing them into objects and their relationships. This paper contributes to the establishment of a broader understanding of the notion of object in conceptual modeling based on *things* that are simultaneously *machines*. In this study, we explored the underlying hypothesis of conceptual models (e.g., UML) to enhance their ontological analysis by using the thing/machine (TM) model, which presents the domain (the subject of description) as thimacs (*thi*ng/*mac*hine-s). Following the philosophical distinction between things and objects, we can specify modeling at two levels: (i) the thinging stage, in which the domain consists of thimacs that emerge from the world, and (ii) the objectification stage, which involves recognizing some thimacs with "hard wholeness," called *objects*. Objects are thimacs that control the handleablity (create, process, release, transfer and receive) of their *sub-parts* when interacting with the outside of the object (analogous to the body parts holding together in an assemblage when interacting with the outside). The results promise a more refined modeling process to develop a high-level description of the involved domain.

*Index Terms – Conceptual model, object orientation, Unified Modeling Language (UML), thing, object, ontology*

## I. Introduction

The object-oriented (OO) approach originated as a programming and software design discipline offering advantages such as reusability, extendibility and portability [1]. In this line of thinking (1993), "a programming language is a matter of introducing *a way of thinking* that is associated with the underlying conceptual framework" [2] (italics added). Writing software in terms of objects results in code is a much more *natural and faithful description* of the problem's structure [3]. According to Madsen [2], object orientation is based on concepts that we first learned in kindergarten: objects and attributes, classes and members and wholes and parts.

---

The Object Management Group (OMG), since 1997, has adopted the Unified Modeling Language (UML) OO methodology as a suitable framework for modeling software-intensive systems. Over the years, the real payoff of the OO approach comes from addressing conceptual issues, rather than implementation issues [1]. The OO paradigm is currently the most popular way of analyzing, designing and developing application systems [4]. Object orientation implies that the notion of *object* plays a central role in modeling and all other notions of a model are viewed by decomposing them into objects and object relationships [4]. Underlying OO analysis and design is the idea that objects in a program can model objects in the real world [5].

### A. About Objects

Objects are studied in many fields of study, and such study concerns the central question of what they *are*. The notion has proliferated across disciplines, which makes its review challenging and allows it to follow multiple pathways and dissimilar foci. Nevertheless, it is necessary to have a coarse awareness of the diversity of the meaning of *object* in various backgrounds.

Ordinarily, upon hearing the word *object*, the first thing we think is that objects are *fixed, stable and unchanging* and therefore contrast with events and processes [6]. Giving a precise characterization of ordinary objects is no easy task. Very roughly, ordinary objects are objects belonging to kinds that we are naturally inclined to regard as having instances: *dog*, *tree*, *table*, and so forth [7]. Some views about the natures of objects may seem at odds with common sense, such as the view that ordinary objects can't survive the loss of any of their parts [7]. According to Tsang et al [8], an object is a *self-contained entity with well-defined characteristics* (attributes) and behaviors, for example, the real-life environment consists of objects such as schools, students, teachers and courses which are related in one way or another. However, *these things are not objects*. Korman [7] observed that *object* is used in its narrow sense, which applies only to *individual* material objects and not to other sorts of entities, such as numbers or events. On the other hand, according to Charles Peirce, "an object is anything that we can think, i.e. anything we can talk about."

In cognitive science, research has suggested that humans recognize objects as *integrated regions that move together and have a hierarchical structure* [9]. According to Booch [10], "you *can do* things to the object, and *it can do* things to other objects, as well" (italics added). As we will see later in this paper, we can interpret this statement as "An object is handled (doing specific five actions) as a whole, and it handles other objects."

According to Capretz [11], the term *object* emerged almost independently in various areas of computer science, e.g., simulation, operating systems, data abstraction and artificial intelligence. Philosophers are always curios about whether a category exists under which *everything* falls. They have come up with many candidates, including *thing*, *being*, *entity*, *item*, *existent* and especially *object* [12]. In ontology-based studies, "objects are *representations* of things" [13]. According to Rumbaugh [14], "Objects serve two purposes: They promote understanding of the real world and provide a practical basis for computer implementation."

In general, object-hood appears to be one of the elusive topics in many fields of study, e.g., philosophy [15]. According to Arnold [15], the word *object* has no "truly standard designation" and because nature does not fix the use or the meaning of the term *object*, we can only look at how people use the term in various ways of thinking and talking. Like other fundamental notions, *object* seems to be primitive, incapable of further conceptual analysis [15]. The property of being an object (object-hood) therefore appears to apply trivially to all conceivable objects. Although we should not expect a strict definition of it, we can still aim for a characterization of the concept(s) of an object, elucidating what objects are without offering criteria for object-hood [15].

In software engineering, Blair [16] pointed out the dualism of an object as a data carrier and something that executes actions: "An object is an encapsulation of a set of operations or methods which can be invoked externally and of a state which remembers the effect of the methods." He defined four dimensions of OO systems: encapsulation (grouping together of various properties), classification (group-associated objects according to common properties), polymorphism (objects can belong to more than one classification) and interpretation.

*B.  This paper*

This paper contributes to the establishment of a conceptual clarity about the notion of *object* as one of the most fundamental terms in software engineering, utilizing a model called a thinging machine (TM) [17-18]. Even though the TM model is supposed to concern conceptual modeling in software and system engineering, this paper concentrates on object-related analysis in the field of general ontology of reality. Such a venture is similar to many studies that have enhanced modeling through ontological analysis. In this context, the underlying paradigm of UML models with respect to ontologies has been explored to provide a comprehensive overview of both domains [19]. Guizzardi, Wagner and Herre [20] evaluated the ontological correctness of a conceptual UML-class model and developed guidelines for how the constructs of the UML should be used in conceptual modeling and ontology representation. Khan and Porres [21] proposed analyzing the consistency and satisfiability of UML models containing multiple classes using the Web Ontology Language. These works are examples of many that enhance conceptual modeling through ontological analysis.

The aforementioned TM model uses the so-called thimacs (*thi*ng/*mac*hine) as building blocks for describing the domain. A thimac is a thing and simultaneously a machine. The machine takes five generic actions: create, process, release, transfer and receive (See Fig. 1). So far, the TM has been utilized to clarify many modeling notions, such as action, activity, behavior, static description and events. Our goal is to define the notion of *object*. We promote the thesis that thimacs in the TM description can be specified at two levels.

(i) *The thinging stage*, at which the (world) domain consists of thimacs that emerge or are thrown toward us (things that emerge from the world). This world of thimacs is—to use Heidegger's words—always "there." The thimacs have a minimum level of organization with respect to the whole-part relationship and their behavior in terms of the five generic actions.

(ii) *The objectification* stage, which involves superimposing on the inside-ness of a thimac a "hard wholeness" to convert it to objects. Objects are characterized by the whole, which takes control of the parts' handleablity (create, process, release, transfer and receive) when interacting with the outside of the object.

We make a new contribution in this paper by binding the notion of *object* to thimacs.

In section 2, we briefly review the basics of TM modeling to make the paper somewhat self-contained. To demonstrate how a TM models domains, in section 3, we introduce two examples: (i) UML class diagram of customers and accounts with a withdrawal operation and (ii) UML relationships to be treated as entities, or what is called "reification." In section 4, we introduce our main thesis, that things are predecessors of objects. In section 5, we discuss the whole-part relationship because it is present in practically all OO modeling languages. Section 6 presents another example of how to represent objects in a TM.

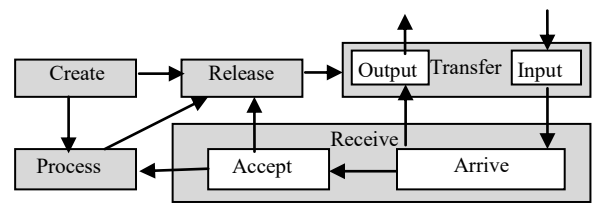

Fig. 1. Thinging machine.

## II.  THINGING-MACHINE MODEL

In the TM model, a thimac generally refers to any definite being. The configuration of thimacs is formed from a juxtaposition of subthimacs that are bonded onto a structure at a higher level, of which they become parts. Note that in this model, the whole (i.e., the thimac) is not identical to the parts that compose it (e.g., the whole has its own machine).

A related notion to the thimac is *object*, which is a special type of thimac, as we will discuss in this paper. Thimacs contain other thimacs (called subthimacs); therefore, a thimac is an assemblage of subthimacs. These assemblages go all the way down and all the way up. In this paper, we claim that objects appear in the model description when some of the thimacs manifest their individuality and independence (as identified by the modeler) in a way that minimizes outside *handling* of their parts (subthings). These thimacs are identified as *objects*. In an object, the parts hold together in an assemblage with the object, which is their agent in the interaction with the outside. Therefore, an object is a thimac with more *organization* that limits flows of actions of its parts.

This two-level ontology has been proposed by many thinkers who assert the existence of something like a pre-objectal or pre-individual background. Deleuze proposed a concept of becoming that entails passage from a *pre-individuality* to an *individuated one* [22]. Sarti, Citti and Piotrowski [], based on Deleuze and Guattari's ideas, proposed a framework "to envisage the emergence of singular forms from the assemblages of heterogeneous operators." According to our interpretation, the difference in forms (e.g., loose thimac vs. firm object) originated from constraining the generic actions in the machine side of the thimac.

The TM model has been developed for several years (see recent paper and former paper in [17-18]). A thimac consists of parts (subthimacs) and generic actions on the whole and on the parts themselves. A thimac exhibits a multiple assembly and has no isolated parts that the actions can reach. A thimac, in contrast to an object, is a loose assembly that can be left from within; that is, there are inner actions, which lead to outside the thimac.

The generic/elementary (having no more primitive action) actions can be described as follows.

**Arrive**: A thing moves to a machine.
**Accept**: A thing enters the machine. For simplification, we assume that all arriving things are accepted; therefore, we can combine *arrive* and *accept* stages into one stage: the **receive** stage.
**Release**: A thing is ready for transfer outside the machine.
**Process**: A thing is changed, handled and examined, but no new thing results.
**Create**: A new thing "comes into being" (is found/manifested) in the machine and is realized from the moment it arises (emergence) in a thimac. Things come into being in the model by "being found." Whereas the action *create* indicates *there is such a thing*. Note that, for simplicity's sake, we omit *create* in some diagrams under the assumption the box of the thimac implies its existence
**Transfer**: A thing is input into or output from a machine.

Additionally, the TM model includes the **triggering** mechanism (denoted by a ***dashed arrow*** in this article's figures), which initiates a (non-sequential) flow from one machine to another. Multiple machines can interact with each other through the movement of things or through triggering. Triggering is a transformation from movement of one thing to movement of a different thing.

### III. TM-MODELING EXAMPLES

In this section, we illustrate the TM modeling using two examples. The first example concerns UML-class diagrams and the second illustrates relationships.

#### A. Example 1 - Bank operations

According to Aoumeur and Saake [23], system designers face challenging problems to keep UML diagrams coherently related. They gave a simple example of an overview of various UML diagrams, namely class diagram, object diagram and associated OCL descriptions. Fig. 2 shows the sample classes in their example. Fig. 3 shows the TM static description of the customer class. We will ignore *create* and the *close* methods in Fig. 2 and focus on the *change address* operation to illustrate the TM model. We assume that address change involves processing the old and new addresses, e.g., storing the old address for future reference.

#### *Change the address*

Every box in Fig. 3 is a machine, including the nameless machine that updates the address. Note that the *create* and *process* of *customer* at the right top corner represent "global" actions, i.e., handling the whole customer as one block. This will be a main feature of objects, as we will discuss later. The address is handled at this global level and not inside the subthimac address. Thus, in the object context, subthimacs are "disconnected" from the outside except through the containing object.

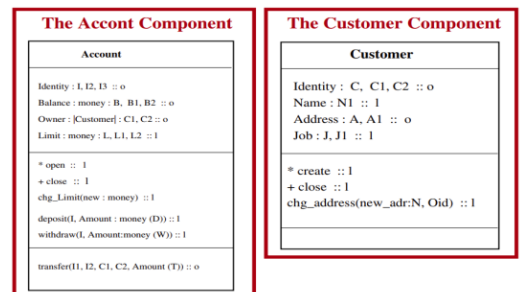

Fig. 2 Classes in the given example.

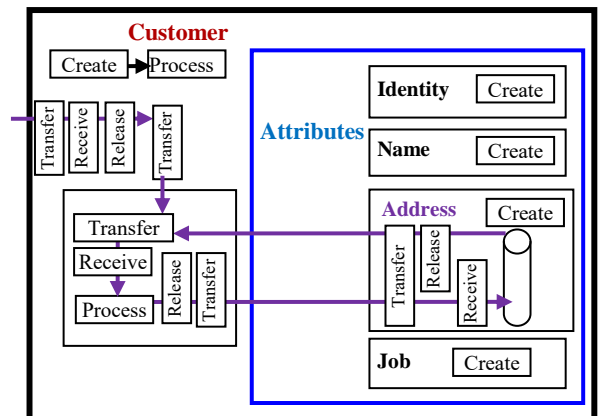

Fig. 3 The static TM model of the customer component

Fig. 4 shows the events TM model. In general, the conception of *event* in a TM follows Whitehead's depiction of events, in which the event constitutes a fundamental datum of reality in space-time, called actual entities/occasions. They are the "final real things of which the world is made up. There is no going behind actual entities to find anything more real" ([24], p. 23). An event in a TM involves a subdiagram of the static description, called a region of the event, plus a time machine [18]. For simplicity's sake, in Fig. 4, events are represented by their regions as follows:

E1: The customer machine is activated, i.e. processed.
E2: The new address flows to the customer machine from the outside.
E3: The old address is retrieved.
E4: The two addresses are processed.
E5: The new address is saved in Address.

Fig. 5 shows the behavior of this change address operation.

*The withdrawal operation*

Fig. 6 shows the *withdrawal* operation in Aoumeur and Saake's [23] example, which involves a *customer* and *account* (pink numbers 1 (customer) and 2 (account) in the figure). We omitted some details inside *customer* and *account* to focus on the withdrawal operation.
- When a customer requests a withdrawal (3), two things are needed.
  • *The identity of the customer* (4), stored in Customer.

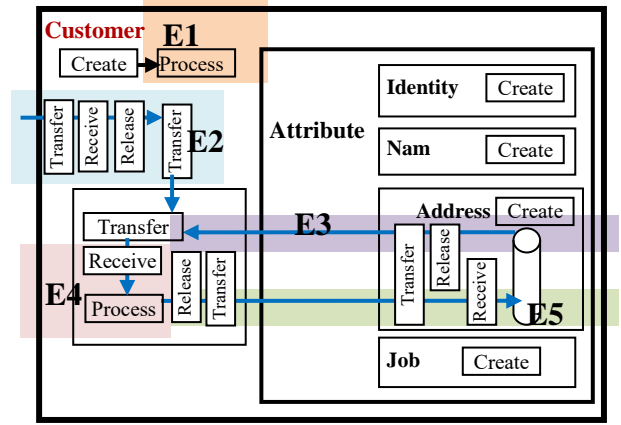

Fig. 4 The events model of the change the address.

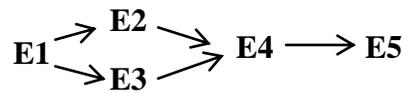

Fig. 5 The behaviour model of the address change.

Here, we assume an already performed login process, which includes accessing the account.
• *An input from the outside* (e.g., from a person) that specifies the amount to be withdrawn (see 5 in Fig. 6).

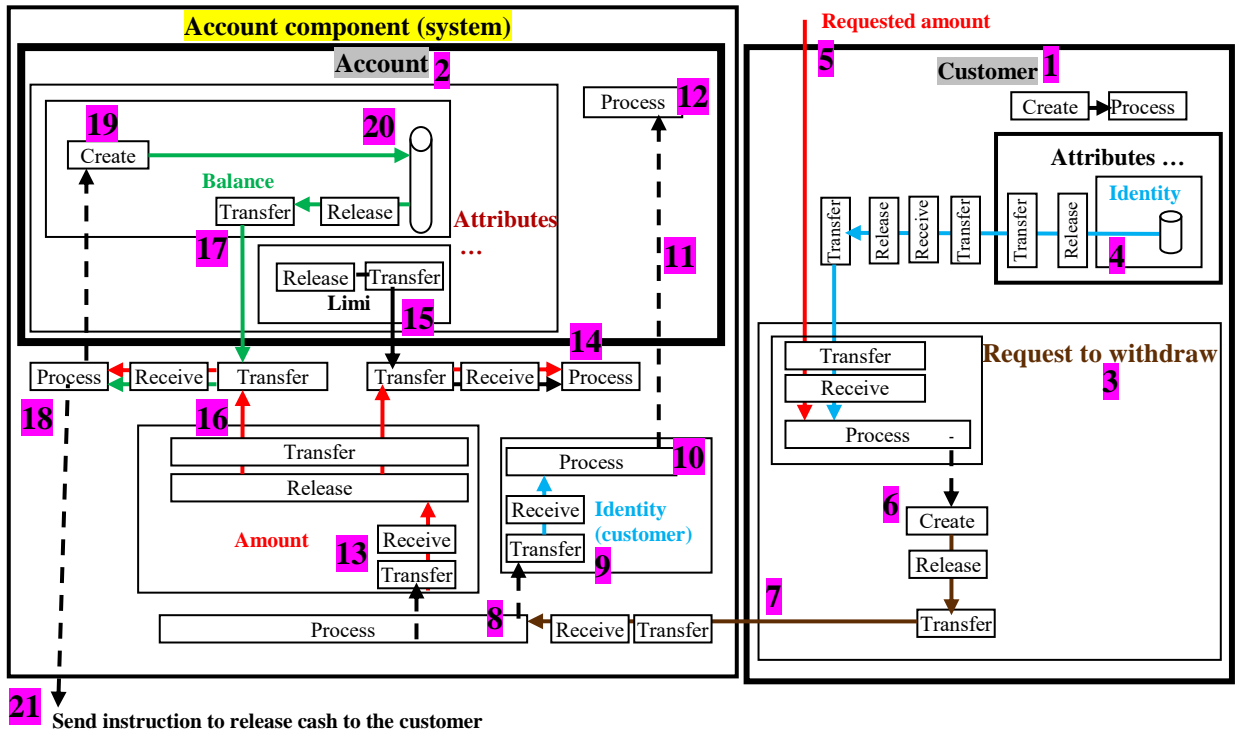

Fig. 6 The static TM model of the withdrawal operation.

- Accordingly, the request for withdrawal with the appropriate data is constructed (6) and flows to the account component (7).
- In the account component, the request is processed (8) to extract the customer identity (9). Note that this extraction is represented by *transfer* and *release*. This process is analogous to the situation of a bus's arrival (a *request in the example*), which indicates a passenger's arrival (*identity in the example*).
- In the account component, the processing of the customer's identity (10) triggers (11) the retrieval and the activation (12) of the customer's account (2).
- Therefore, the amount (13) is extracted from the request for withdrawal, to be compared (14) to the limit (15). For simplicity's sake, we ignored the cases of error here, assuming every process is accurate.
- Additionally, the extracted requested amount (17) and the balance (7) are processed (18) to create the new balance (19 and 20). Instructions are also sent to release cash to the customer (21), i.e., ATM cash.

To save space, we do not list the description of events; instead, Fig. 7 shows the events model and Fig. 8 shows the behavior model of the withdrawal operation.

*B. Example 2 - Reification*

According to Halpin [25], some modeling approaches allow instances of relationships to be treated as entities in their own right.

In UML, this process is called "reification." In practice, objectification is prone to misuse, and some modeling approaches provide incomplete or flawed support for it [25]

Fig. 9 displays a simple model in which *country* is depicted as playing sport, which is objectified as the object type *Playing*. Associations are connected to the classes whose object instances can play their roles. UML encodes facts using either associations or attributes. It treats the association class *playing* as identical to the association [25]. Fig. 10 shows the TM static representation of this situation. Fig. 11 shows the events that construct the relationship between country and sports. First (it is assumed that country and sport exist), *country* and *sports* flow to the construct *country-sport*, and then the rank is added.

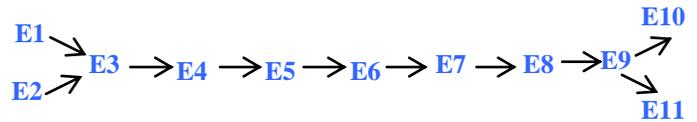

Fig. 8 The behaviour model of the withdrawal operation.

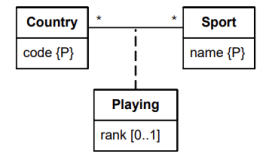

Fig. 9 Objectification of Country plays Sport as Playing in UML (From [25]).

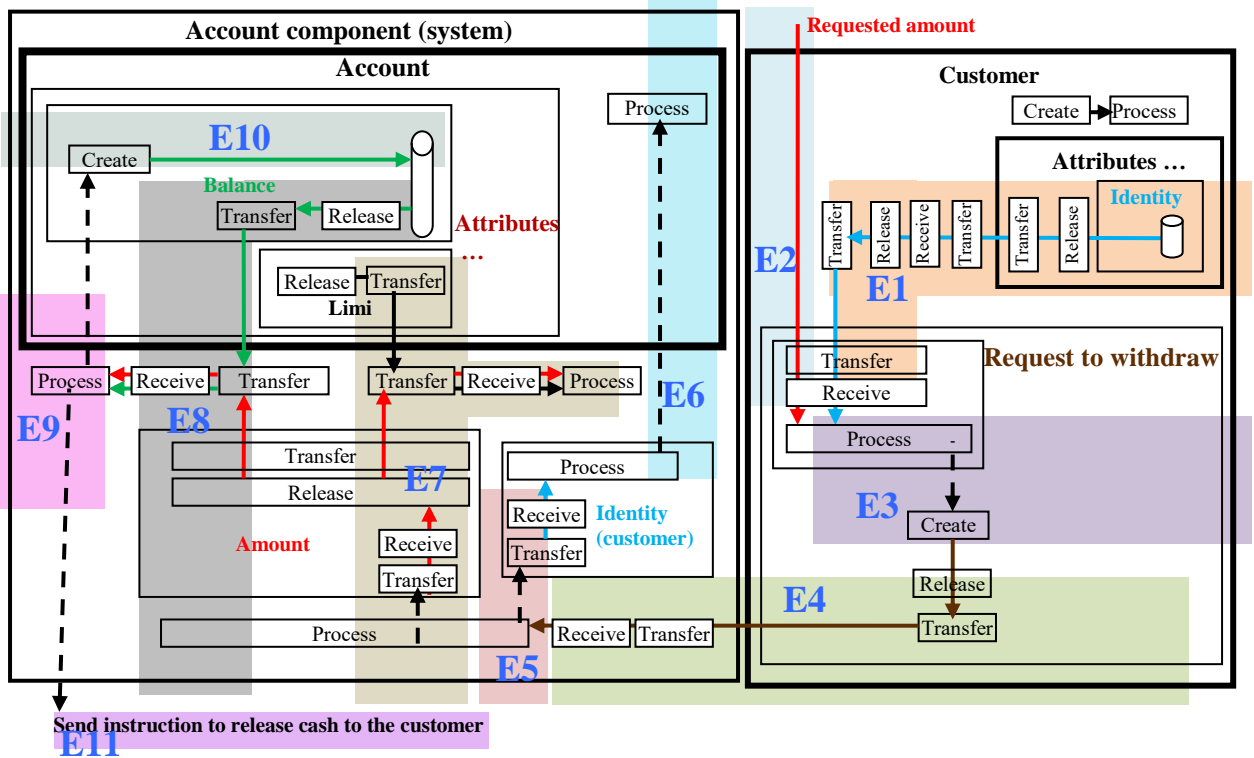

Fig. 7 The events model of the withdrawal operation

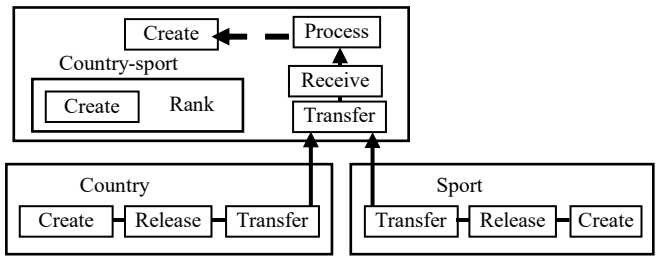

Fig. 10 The TM representation of Halpin's objectification.

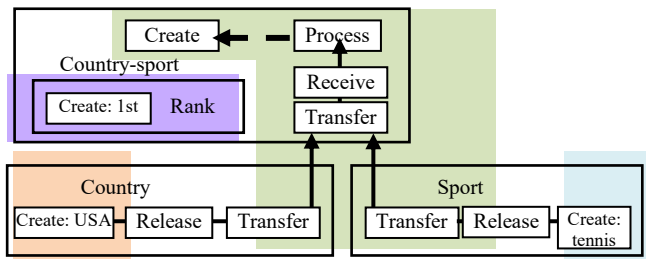

Fig. 11 The object (USA, tennis, 1st).

## IV. BASIC THESIS OF THIS PAPER: *THINGS* ARE PREDECESSORS OF *OBJECTS*

In the remainder of this paper, we identify the role of the notion of *object* in the TM representation. This identification requires distinguishing between things and objects. The discussion involves how to complement the TM description with objects and includes class examples from UML. It is important to note that the alignment of the OO TM with UML is not an issue here and that because of the extensive details of the UML construct, such an alignment will be studied in further works.

According to Bodei and Baca [26], "there is a misunderstanding resulting from the lack of distinction between 'thing' and 'object,' words that have become confused over time, causing a series of cascading misinterpretations that blur both philosophical thinking and common sense." Šerpytytė [27] stated that the problem of the difference between things and objects is one of the most decisive issues for the conception of the real:

> These words are usually used interchangeably – and not only in their everyday usage. There are some contemporary philosophical positions that consider almost "every*thing*" as an object; on the other hand, there are proponents of a strict separation of objects and things. How did it happen that the concept of thing (*res*) and object (*obiectum*) not only began to theoretically "compete" with each other but also sometimes came to represent differently conceived realities, and even occasionally came to represent an identical conception of reality?

In OO ontology [28], objects "swallow up" things; on the other hand, there are proponents for a strict separation between objects and things, e.g., Martin Heidegger [27].

### A. An approach to positioning objects in TM Modeling

In the following, we selectively (probably unfairly) adopt Šerpytytė's [27] writings to support the idea in this paper that things (thimacs) are predecessors of objects in conceptual modeling, as Fig. 12 shows. Accordingly, to start with things, one traditional meaning of a thing is something from which "the totality of things, reality *originates* and relates to" [27]. On the other hand, some medieval scholars used the term object (objectum) to mean "to throw in front," "to put before" or disclosed for the subject as something before him [27]. The object is conceived as the sensibly *graspable thingliness* and a reality "assembled" from tangible things [27].

In this paper, we capitalize on the definition of *thimac* as a thing and a machine. Accordingly, the machine as the source of actions is utilized to limit the scopes of these actions at the levels of the whole and parts of the thimac. The following example illustrates the basic idea of restraining the actions.

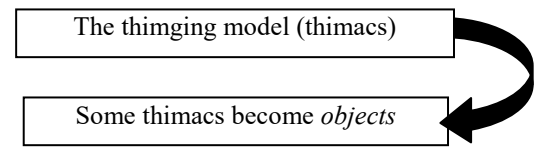

Fig. 12 The object-orientation approach involves identifying objects in the TM descriptions. Objects are thimacs that dominates the interactions of its parts with the outside of the object.

### B. Example: A non-object thimac vs. an object

Consider the following situation described by Wallace [29].

> Suppose your chair is composed of only two parts: a seat and a (thick) leg. Here is a seemingly simple question: how many *objects* are there? You might be inclined to say one – there is just one chair, after all. But I said that the chair was composed of a seat and a leg. So it seems we have to count the seat and the leg, too. So are there three *things*? The chair is a material *object* that occupies region. The seat and the leg are material *objects* that occupy region. This is a case of complete spatial overlap. Our seemingly simple universe containing an ordinary *object* and its parts has revealed itself to be quite complicated indeed; we cannot even say how many *objects* there are!

In TM modeling, the chair is conceptualized at two levels. In the thinging stage, the chair and its parts are as shown in Fig. 13. For simplicity's sake, we consider only movement actions to the outside. In Fig. 13, the chair is initially a thimac that is constructed (as constructor in OO programming) as a whole and includes its subthimacs, seat and leg, which can move to the outside freely (i.e., not tied to the whole chair). In Fig. 14, the chair is an object.

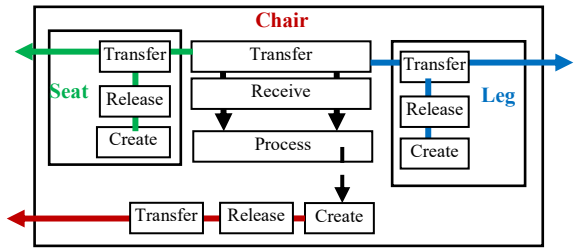

Fig. 13 The chair as non-object thimac.

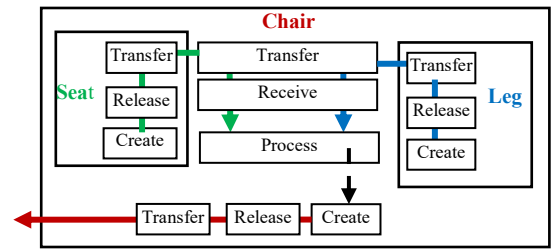

Fig. 14 The chair as non-object.

The new view in Fig. 14 is that the seat and the leg move in the context of the whole chair. Therefore, in the case of the object, the movement of the parts is bound to the movement of the whole chair. This restriction of the parts' freedom of movement is facilitated by the notion of a machine as an integral constituent of the thimac. When the chair is an object accessing, say, the individual seat requires bringing the chair and processing it (e.g., dismantling it).

Accordingly, things have more freedom of handleability than objects and our first encounter in the world is with thimacs, not objects. We encounter thimacs and then generate objects from some of the thimacs that we come across with them. We view the thimac as the "nature itself diversified by means of something" [30]. It lacks a single center of regulation, and its parts are distinct free units. We conceptualize things as being pre-given to the mind. Some of these things become objects when they can be grasped by mental acts (it *can* be represented as a handleable whole of parts). Note that part of this description uses selected ideas from Meinong [31].

The non-object thing has unity, but it is accidentally bonded, as a flock of birds flying together. At any time, one or some birds may fly away. An object is a thimac concentrated as one definite whole that dominates the actions of its parts, especially with the outside of the object. The parts depend on their containing object for outside interaction. An object is a unity of composition. Here, *unity* depends on the object's handleablity (create, process, release, transfer and receive). Therefore, the uniting bond is the centralized handleabilty. The object unity seems to be a Gestaltic unity, in which the way in which each part is handled is derived from the structure of the whole.

The notion of *class* can be applied to non-object things and objects as a mind-based specification of types of instances of things that have a common structure. An object/class reflects its objects' unity, and each object reflects the unity of its parts' handleability.

Each set of objects in an object/class has no part that communicates or interacts with the outside without the acknowledgement and control of its object/class. The uniting bond between separate objects of that class is the standard handleability specified in the class. The *create* action generates one object, which *implies* the creation (internally) of various parts of the object. *Process* entails processing the whole object to access its parts. *Release*, *transfer* and *receive* are object actions. Accordingly, the object has interiority, which includes its parts as thimacs that form the object's "life" (interior static description, events and behavior). The object interacts with other outside objects, and control flows from this interior life to/from the outside. The general TM model does not impose such restrictions on parts of thimacs. To superimpose the object-orientation feature, the modeler has to specify those things that are objects.

*C. TM Presuppositions*

Thimacs are recognized at the pre-objectification stage (thinging stage), at which thimacs "are there." From such a base, *objects* are identified. Accordingly, the following statements are prerequisites for TM modeling:

- Everything in the pre-objectification stage is a *pre-object* thimac (thing/machine). The universe is constructed from a set of thimacs, which exhibit "togetherness" to form a bounded whole. The assumption here is that "the world consists of some fixed totality of mind-independent objects" [32] and that "...the world, and not thinkers, sorts things into kinds" [32]. The flow (create, process, etc.) occurs, in some sense, *prior* to the things that form and move in this flow. This situation can be projected onto an analogous situation that Bohm [33] described:

  > One can perhaps illustrate what is meant here by considering the "streams of consciousness." This flux … is evidently prior to the definable forms of thoughts and ideas which can be seen to form and dissolve in the flux, like ripples, waves and vortices in a flowing stream. As happens with such patterns of movement in a stream some thoughts recur and persist in a *more or less stable way*, while others are evanescent. (italics added)

  The various patterns that can be abstracted from the initial depiction have a certain *relative autonomy and stability* that the thimacs capture. The thimac is a universal indication of any form of existence (creation), change (process) or movement (release, transfer and receive).

- Objectification is a mind-dependent processing of the pre-objectification stage, which ignites the recognition that some thimacs have "hard wholeness" with respect to their subparts such that they have to be handled wholly (together). These are non-arbitrary sets of things that have *natural conceptual wholes* (a term taken from Guizzardi [34]).

- Objectification has utilized pre-objectification to harden the thimacs' fuzziness state. It created objects by selecting thimacs as "wholly handled objects"; things can be created, processed, released, transferred or received only as a whole. If you want to handle part of it, you must process the whole, and then you can handle the part.

Note that the OO approach applies the *everything-is-an-object* principle to the world [36].

## V. WHOLE/PART THIMACS

*Parthood* is a relationship of significant importance in conceptual modeling, being present in practically all OO modeling languages even though it is understood only intuitively. Much disagreement remains regarding what characterizes this relationship as well as the properties that part-whole relationships should have from a conceptual point of view [34]. A central issue in this context is what makes something an *integral whole* composed of many parts. Another problematic feature is the non-differentiation among the roles that parts play within the structure of an aggregate, e.g., whether things can share parts and whether a thing only exists as part of a specific whole [34]. A great deal of confusion regarding the whole-part relationship comes from current thinking, where static and dynamic levels are mixed up.

### A. The composite relationship

According to Mylopoulos [35], a UML class describes a group of objects with similar properties (attributes), common behaviors (operations) and common relationships to other objects and common (semantics). Classes and objects do not exist in isolation from one another. A relationship entails connections among objects. UML includes multiple types of relationships, e.g., generalization, association, aggregation and composition [35]. A composition relationship implies strong ownership of *the part by the whole*. It also implies that if the whole is removed from the model, so is the part. For example, the relationship between a person and her head is a composition relationship. In a composition relationship, the whole is responsible for the disposition of its parts, i.e., the composite must manage the creation and destruction of its parts. Mylopoulos [35] provided the composite relationship shown in Fig. 15.

### B. TM modeling

Consider Fig. 16, which shows the corresponding TM static model in which the order thing (1) triggers (2), creating OrderItem (3), which includes product (4). Fig. 16 is an initial model, as illustrated previously in the chair example in Fig. 13, where we know there is a chair but we are not sure about its structure (bonding to the seat and the leg). In Fig. 16, the modeler may consider that the *order*, *orderItem* and *product* are objects. These objects are the only handleable things at the global level, as Fig. 17 shows (the objects are dotted boxes), in which confirmation has been added to the description. The model can be modified, e.g., if we want an object of *order* composed of orderItems, we can include that in a dotted box that includes the order and its orderItems. Fig. 17 does not coincide with the compositional semantics of the UML-class diagram in Fig. 15, but it seems more appropriate. For example, it seems that *product* is not a subproject of *orderItem* but has a relationship with it.

Fig. 18 shows the behavior model of creating a new order with its orderItems. First, the new order is constructed (E1) and processed (E2) repeatedly to create its orderItems. Note that in Fig 16, we have not shown the object order's construction. We can do so, as in Fig. 13, in which the chair is constructed with the seat and the leg. Because order is declared as an object, if one wanted to change the *date* of the order, they would have to process the whole object order, extract the date, update it and then create a new order as a replacement for the old order.

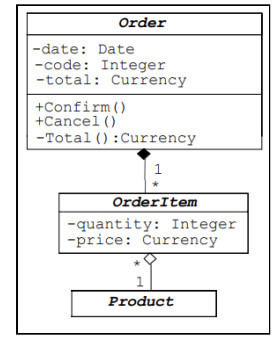

Fig. 15 An example of a composite relationship (From [35]).

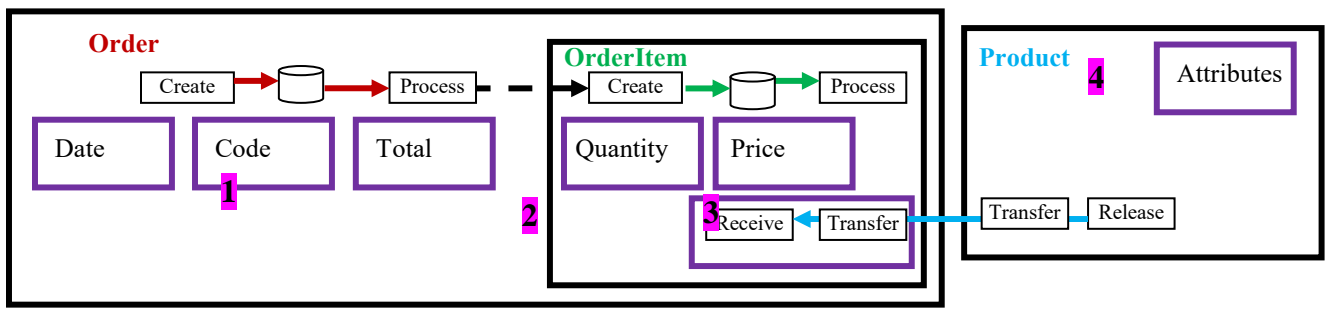

Fig. 16 Pre-categorization picture of the order environment.

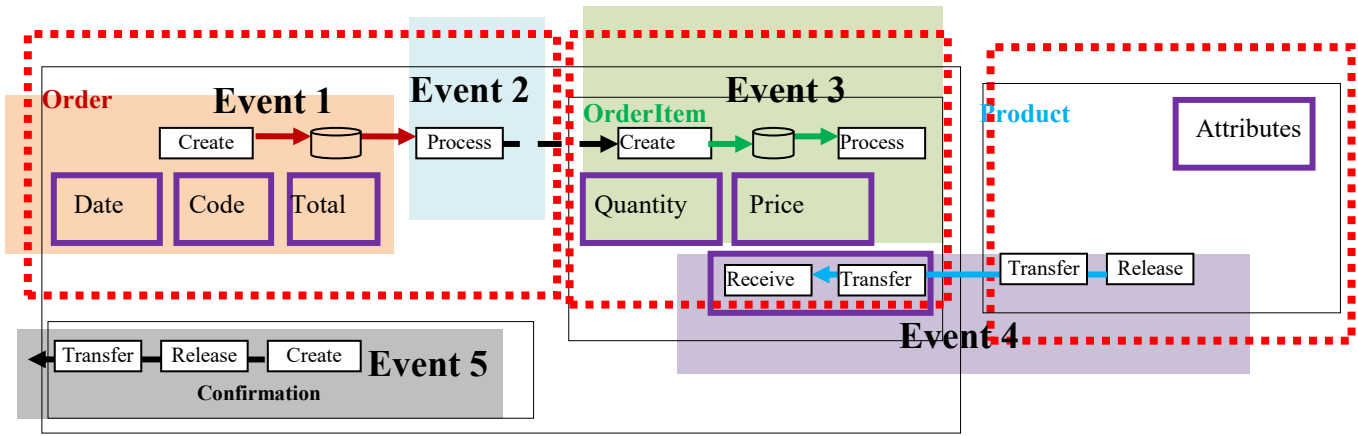

Fig. 17 Objectification of the sample with events and confirmation.

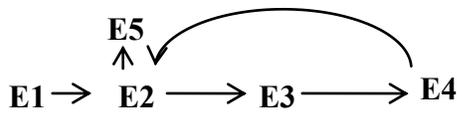

Fig. 18 The behaviour model for creating a new order with its orderItems.

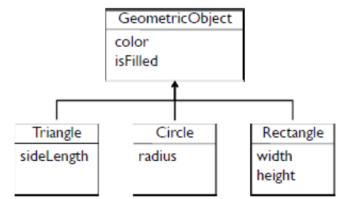

Fig. 19 UML-class diagram (From [37]).

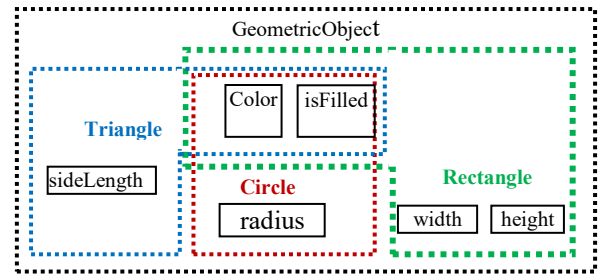

Fig. 20 The four objects in the example.

## VI. OBJECTS AND CONCEPTUAL REPRESENTATION

Although this paper laid the ground for defining objects based on thimacs, many issues related to the OO approach remain to be discussed, especially concerning relations among objects, e.g., subclasses, inheritance and compositions. However, the TM modeling is a purely conceptual approach and can therefore form a foundation for conceptualizing object-based modeling, as the following example shows.

According to [37], suppose that we needed to be able to draw circles, equilateral triangles and rectangles. Each of these shapes has some unique instance variables that would be necessary for drawing them. For example, our Triangle class would need to have a sideLength, the Rectangle class would need to have width and height and our Circle class would need to have a radius. However, they should have some instances in common, perhaps a "color" and an "isFilled." The best thing to do in this situation would be first to create a class, perhaps named "GeometricObject," that would contain the common-instance variables and then create subclasses of GeometricObject, called Triangle, Rectangle and Circle, which would include the additional (unique) instance, as Fig. 19 shows [37].

In the corresponding TM model, the solution involves four objects, as Fig. 20 shows, where the machines of various objects are not shown. Because they are objects, each can perform actions (create, process, etc.) with the outside and in the context of the encompassing object. Fig. 21 shows the construction of GeometricObject, with Triangle, Rectangle and Circle, showing only the rectangle actions.

Nevertheless, such an OO approach seems to complicate the conceptual picture for the purpose of, possibly, saving storage. Fig. 22 presents a simpler conceptual model in which GeometricObject is modeled as the union of the other three sets of objects.

## VII. CONCLUSION

The TM methodology is based on a speculative scheme focusing on conceptual modeling that views the world in terms of thimacs with a dual nature: thing and machine. The machine has five generic actions. The new aspect in this paper is that the machine side of the thimac is used to restrict the flow, thus developing the notion of *object*.

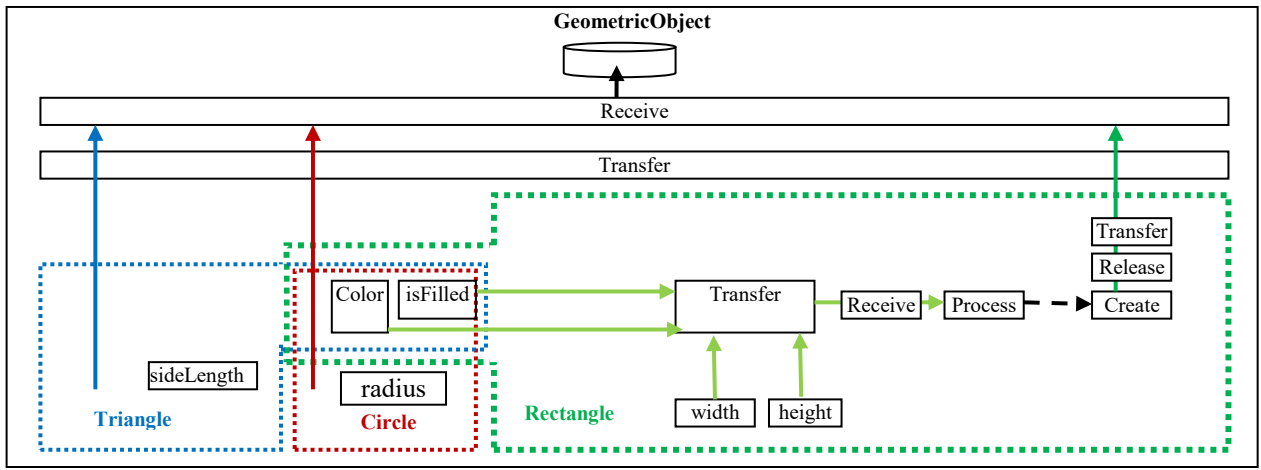

Fig. 21 The construction of Rectangle and GeometricObject.

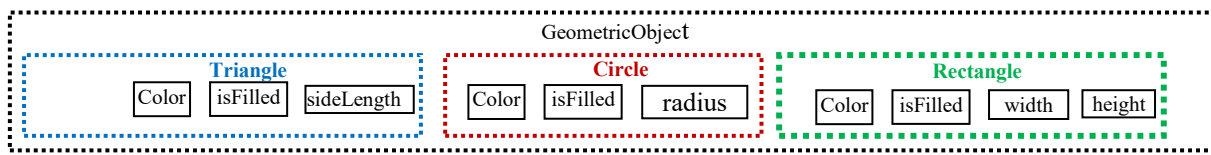

Fig. 22 Conceptual simplification of the four objects in the example.

An object is a thimac such that all actions related to the outside flows are restricted to the whole thimac level. Based on such a concept, the notion of *object* in conceptual modeling is partially explored (e.g., using some UML examples) to substantiate the feasibility of such a definition of *object*. The initial indications point to the viability of this definition of *object*; however, such a venture seems to require further examination, especially in applying it to many OO notions, e.g., UML constructs.